# STRUCTURE AND PROPERTIES OF EPITAXIAL THIN FILMS OF $Bi_2FeCrO_6$: A MULTIFERROIC MATERIAL POSTULATED BY AB-INITIO COMPUTATION

R. Nechache[1], C. Harnagea[1], L.-P. Carignan[2], D. Ménard[2], and A. Pignolet[1]

[1] Prof. A. Pignolet, R. Nechache, C. Harnagea
INRS Énergie, Matériaux et Télécommunications,
1650, boulevard Lionel-Boulet,
Varennes (Québec), J3X 1S2 (Canada).
E-mail : pignolet@emt.inrs.ca

[2] Prof D. Ménard, L-P. Carignan
École Polytechnique de Montréal,
Département de Génie Physique,
P.O. Box 6079, Station. Centre-ville,
Montréal, (Québec), H3C 6A7 (Canada).
E-mail : menard@polymtl.ca

## ABSTRACT

Experimental results on $Bi_2FeCrO_6$ (BFCO) epitaxial films deposited by laser ablation on $SrTiO_3$ substrates are presented. It has been theoretically predicted using first-principles density functional theory that BFCO is ferrimagnetic (with a magnetic moment of $2\mu_B$ per formula unit) and ferroelectric (with a polarization of ~80 $\mu C/cm^2$ at 0K). The crystal structure investigated using X-ray diffraction shows that the films are epitaxial with a high degree of crystallinity. Chemical analysis carried out by X-ray Microanalysis and X-ray Photoelectron Spectroscopy indicates the correct cationic stoichiometry in the BFCO layer, namely (Bi:Fe:Cr = 2:1:1). Cross-section high-resolution transmission electron microscopy images together with selected area electron diffraction confirm the crystalline quality of the epitaxial BFCO films with no identifiable foreign phase or inclusion. The multiferroic character of BFCO is proven by piezoresponse force microscopy (PFM) and magnetic measurements showing that the films exhibit ferroelectric and magnetic hysteresis at room temperature. The local piezoelectric measurements show the presence of ferroelectric domains and their switching at the sub-micron scale.

# INTRODUCTION

Multiferroics are materials which display both ferroelectric and magnetic ordering and are promising candidates for innovative devices such as transducers, sensors [1] and memories for data storage. Recently, using first-principles density functional theory, Baettig and Spaldin [2] studied $Bi_2FeCrO_6$ (BFCO) and predicted ferrimagnetic and ferroelectric properties allegedly much larger than those of the known multiferroic materials. They theoretically predicted that at 0 K BFCO would be both ferrimagnetic and ferroelectric, with a magnetic moment of $2\mu_B$/f.u. (formula unit) and a polarization of 80 $\mu C/cm^2$, respectively. The predicted ground state structure is very similar to that of the rhombohedral R3c structure of $BiFeO_3$ (BFO), but with $Fe^{3+}$ cations substituted with $Cr^{3+}$ cations in every second (111) layer, reducing the symmetry to the space group to R3.

The first successful synthesis and the measurement of structural and multiferroic properties of this hypothetical material were recently reported by our group [3]. Here we present a more extensive study of their structural, electric and magnetic characterization. These results are compared to the properties predicted by ab-initio calculations [4].

# EXPERIMENTAL

PLD is one of the most suitable and frequently used techniques to grow heterostructures of complex multi-component oxides in moderate oxygen pressure. Using this process we prepared heterostructures of BFCO on conductive $SrRuO_3$ (SRO) films on (100) oriented $SrTiO_3$ (STO) substrates and directly on conductive (100)-oriented niobium doped $SrTiO_3$ (STO:Nb). The epitaxial SRO films were prepared from a stoichiometric ceramic target and were used both as bottom electrode for electrical measurements and to promote the heteroepitaxial growth of BFCO. The epitaxial BFCO films were obtained from a dense ceramic target composed of a 50 mol.% $BiFeO_3$ and 50 mol.% $BiCrO_3$ mixture. The substrate temperature during depositions was 750°C and the BFCO films were deposited at a growth rate of ~1.5 Å/sec in oxygen ambient at a pressure of 20 mTorr. To eliminate possible oxygen vacancies, the films were cooled from 750°C down to 400°C at a slow cooling rate of 8°C/min in oxygen atmosphere, and maintained at this temperature for one hour, prior cooling down to room temperature.

The BFCO film stoichiometry was analyzed by Rutherford Backscattering Spectrometry (RBS) (Van der Graff accelerator 350 keV) as well as using Energy Dispersive X-ray Spectroscopy (EDXS) in a Scanning Electron Microscope (SEM). The oxidation states of the cations, as well as the composition, were also investigated across the whole layer's depth for a 300 nm thick film using X-ray Photoelectron Spectroscopy (XPS) (ESCALAB 220i-XL system). Depth profiling was performed with intermittent $Ar^+$ ion sputtering. The sputtering was done with a 3 keV $Ar^+$ ion beam (beam current: 0.1 - 0.2 µA) in 2-3 ×$10^{-8}$ Torr of Ar. The ion beam was scanned over an area of 2 × 2 $mm^2$. Under the conditions used, the average sputtering rate measured was about 0.033 nm/sec. The X-ray photoelectron Bi 4f, Fe 2p, Cr 2p and O 1s core level spectra of the BFCO successive surface layers were collected with a hemispherical electron energy analyzer using a Mg Kα X-ray twin source (1253.6 eV). The pressure inside the analysis chamber was maintained below $3 \times 10^{-9}$ Torr during XPS measurements. The atomic concentration was determined by using the sensitivity factors supplied by the instrument manufacturer. The binding energy scale was corrected and calibrated using the Bi $4f_{7/2}$ peak position (158.8 eV). The crystal structure of the BFCO films, as well as the quality of their epitaxy were investigated using X-Ray Diffraction (XRD) (PANalytical X'Pert MRD 4-circle diffractometer).

Local piezoelectric measurements were carried out using piezoresponse force microscopy (PFM) [5,6,7]. Here we used a DI-Enviroscope AFM (Veeco) equipped with a NSC36a (Micromasch) cantilever and tips coated with Co/Cr. We applied an ac voltage of 0.5V at 26 kHz between the conductive tip and the SRO layer located beneath BFCO and we detected the BFCO film surface induced piezoelectric vibrations using a lock-in amplifier from Signal Recovery (model 7265). Local hysteresis measurements as well as poling over micron-sized areas were achieved using a DC-source (Keithley 2400) to apply bias voltages to the bottom electrode of the sample.

The magnetic hysteresis (M-H) loops were measured at room temperature using a Vibrating-Sample Magnetometer (VSM) (Model EV9 from ADE Technologies). A maximum saturating field of 10 kOe was initially applied parallel to the film plane, then decreased down to -10 kOe in steps of 500 Oe, and back up to 10 kOe. The field steps were decreased down to 50 Oe around the zero field range for better resolution. An average factor of 20 measurements per field points provided an absolute sensitivity of about $10^{-6}$ emu. The observed hysteresis clearly indicated the presence of an ordered magnetic phase. The magnetic responses of the sample holding rods and of bare substrates (without BFCO) were also measured and subtracted in order to ensure that the ferro- or ferrimagnetic signal was originating solely from the BFCO films.

## RESULTS AND DISCUSSION

### Chemical and Phase Analysis

RBS analysis of the BFCO/SRO/STO (100) heterostructure (spectrum not shown) indicates that the cationic ratio Bi/(Fe+Cr) was close to unity (within experimental resolution of 3%). The Fe/Cr ratio and the oxygen content are difficult to determine due to the close position of the Fe and Cr peaks and the small oxygen scattering cross-section, respectively. However, this information can be found using other chemical analysis techniques, such as X-ray microanalysis inside a Scanning Electron Microscope (SEM) and X-ray Photoelectron Spectroscopy (XPS) analysis. Both techniques confirmed that the average Fe/Cr ratio is about unity (Fig.1(a)), within the error of the techniques of a few percent. The Fe and Cr oxidation states were investigated for BFCO/STO sample with x-ray photoelectron spectroscopy. The representative scans of the Fe 2p and Cr 2p are shown in Fig. 1(b). The position of Fe $2p_{3/2}$ line is expected to be 711 eV for $Fe^{3+}$ and 709.5 eV for $Fe^{2+}$, and the position of the satellite peak is expected at 719 eV for $Fe^{3+}$ and 716 eV for $Fe^{2+}$. [8] From the spectrum of Fig. 1(b), we deduce that the oxidation state of Fe in BFCO thin films is $Fe^{3+}$ and that there is no evidence for the presence of $Fe^{2+}$ within a resolution of a few atomic percent. XPS depth profiling results [9] reveal that the oxidation state of Fe and Cr ions in the film remains 3+ throughout the films thickness and hence exclude (within the detection limit of the technique) the presence of magnetite ($Fe_3O_4$) iron-oxide impurities in our BFCO films. In another hand, an EDXS line profile obtained from cross-sectional TEM image of BFCO/SRO/STO100 heterostructure (not shown here) reveal that the concentration of Bi, Fe and Cr cations are nearly constant across the film thickness, approximately 20%, 10%, and 10 %, respectively (within the experimental error of 3 at%), the expected values for BFCO. This result confirms that Fe and Cr ions are present in the right ratio homogeneously distributed throughout the depth.

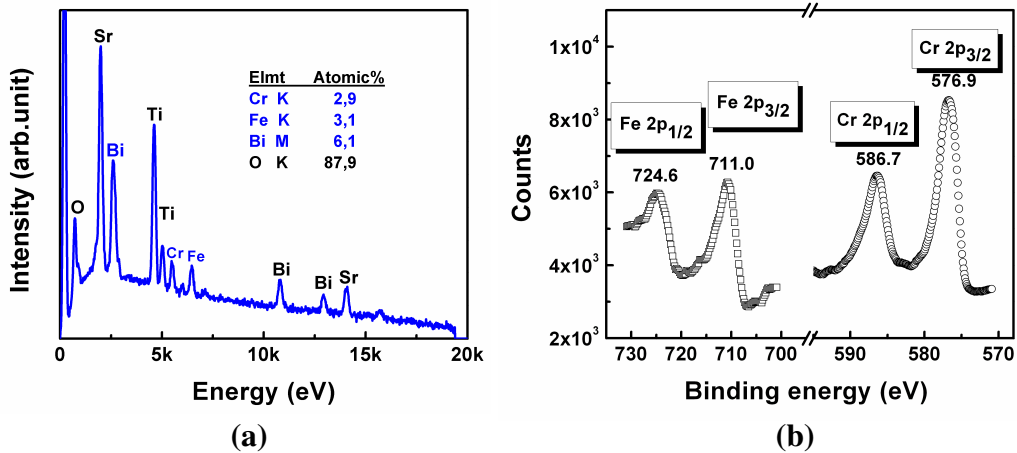

***Figure 1.*** (a) X-ray microanalysis spectrum obtained from a 250 nm thick BFCO film deposited directly on a Nb doped-STO (100) substrate. (b) X-ray photoelectron spectrum of the Fe 2p and Cr 2p lines for 250 nm thick BFCO on STO.

Figure 2(a) shows the XRD θ-2θ scan of a 150 nm thick BFCO film typical of films grown on epitaxial SRO buffer layers on (100)-oriented STO substrates. The BFCO films are epitaxial and no reflections that would be indicative of any secondary phase are observed. The represented θ/2θ scan is showing the 00*l* peaks of BFCO near the *h*00 reflections of STO. The calculated out-of-plane lattice parameters are typically ~3.95 Å. XRD analysis with extremely long counting times in the range 2θ= 91° to 97° were also performed in order to exclude the presence of magnetic parasite phases, as those reported by Bea *et al*..[10]. No peaks corresponding to α- or γ-$Fe_2O_3$ reflections were detected, even after these extra long scans.

In the spectrum of Fig. 2(a), the 00*l* reflection of the 30 nm thin SRO layer are not visible since they are located at angles very close to that of the BFCO 00*l* reflections, and thus are buried within the 00*l* reflections of the upper layer. Such a high value of the out-of-plane lattice parameter of the 30 nm thick epitaxial SRO bottom electrode suggests a fully strained epitaxial SRO layer. The in-plane orientation was investigated by Φ-scans, using the pseudo-cubic {101} reflections of both BFCO and STO (Fig. 2(b)). The fourfold symmetry and relative position of the STO and BFCO peaks indicate a "cube-on-cube" epitaxy of BFCO on SRO/STO.

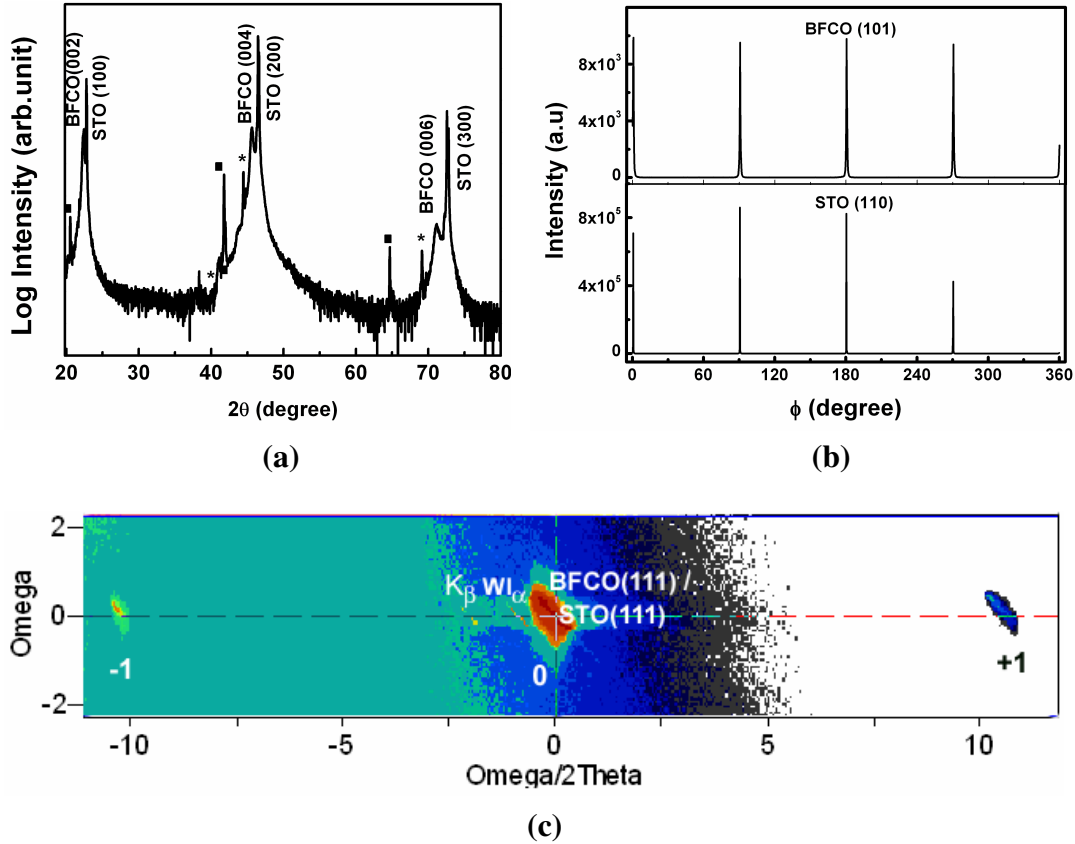

*Figure 2.* **(a)** XRD Pattern for a 150 nm-thick films grown on SRO on STO (100) and **(b)** Φ-scan of the (101) plan. Squares and stars represent the $K_\beta$ lines of STO and $WL_\alpha$ X-ray tube contamination, respectively. **(c)** Reciprocal space mapping around the BFO and STO 111 reflection showing the superstructure related to the Fe-Cr ordering within the BFCO unit cell.

In order to investigate more in details the crystallographic structure of the BFCO film, we performed X-ray Reciprocal Space Mappings (RSM). A reciprocal space map recorded near the BFO and STO reflections is shown in Fig. 2(c). The RSM reveals the presence of two additional superstructure satellite peaks (labelled -1 and +1 in Fig. 2(c)) in addition to the BFCO(111) / STO(111) RLP (reciprocal lattice point). The superlattice period corresponds to 4.6 Å, two times pseudo-cubic $d_{111}^{BFCO} \approx d_{111}^{STO} = 2.3$ Å, and reveals the existence of a Fe/Cr cationic ordering along the [111] crystallographic direction in agreement with the rhombohedral structure obtained by ab initio calculations.[4]

To confirm that the films are homogenous and do not contain significant amounts of secondary phases, we performed a Transmission Electron Microscopy (TEM) characterization. Low-magnification TEM image (Fig. 3(a)), as well as Selected Area Electron Diffraction (SAED) patterns (Fig. 3 (c-d)) obtained from a cross-section of a (001)-oriented BFCO film confirm the high crystalline quality of the BFCO film and do not provide any evidence of second phases or inclusions of any kind. The c-axis of BFCO is found to be parallel to the film normal. Indexing of this SAED pattern with pseudo-cubic indices yields an in-plane lattice parameter very close to that of STO substrate (~3.91 Å) and an out-of-plane parameter of 3.95, Å close to that calculated from the XRD spectrum of this film.

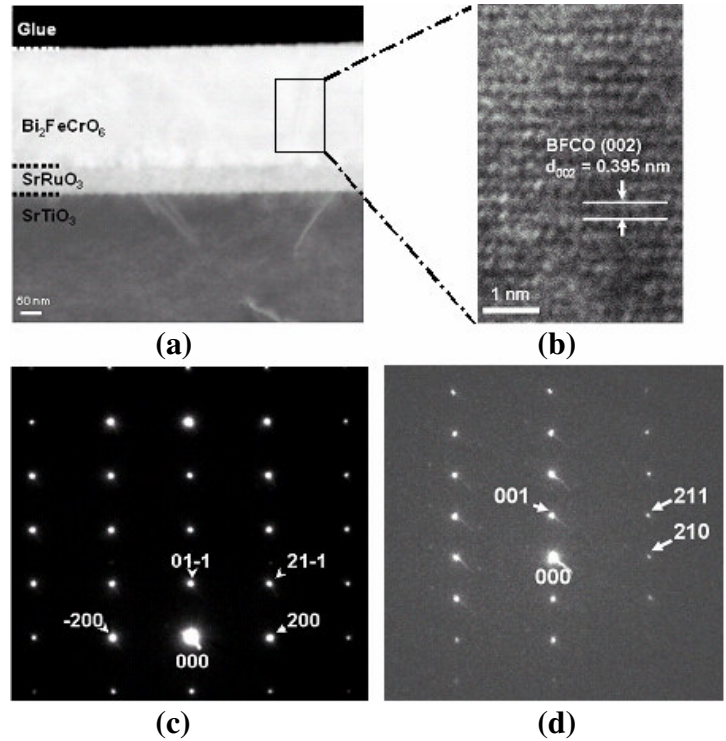

*Figure 3.* **(a)** Cross-section TEM image showing a 300 nm thick epitaxial BFCO layer on an epitaxial SRO layer on STO (001). The direction of observation is [010]. **(b)** High-resolution TEM image of the epitaxial BFCO layer taken along [210] direction. Electron diffraction pattern of the same a 300 nm thick BFCO layer as in (a) and (b) along **(c)** the [011] direction and **(d)** the [1-20] direction.

The same out-of-plane lattice parameter is given by high resolution TEM (HRTEM) image in Fig. 3(b) showing (001) atomic planes. These room temperature lattice parameters are 1%, respectively 2%, larger than those predicted by first-principles density functional theory [4] (the latter being, strictly speaking, computed at 0K), however BFCO lattice parameters are still smaller than those of BFO.

In order to demonstrate the multiferroic character of the films, we measured their electric, electromechanical and magnetic properties.

## Local Ferroelectric Properties

Macroscopic ferroelectric hysteresis loops were similar to those already reported for our BFCO films on SRO buffer layers. [3] They show a relatively low polarization because we could not saturate the hysteresis loops due to a sizable leakage current. We will therefore present here only local electromechanical results.

The surface morphology of the as-deposited BFCO film was examined using atomic force microscopy (AFM). The film had small grain size, typically less than 250 nm. A root-mean-square (rms) roughness of about 2 nm for a 10 × 10 µm² surface of film has been measured.

Ferroelectric characterization was performed at the nanoscale using piezoresponse force microscopy, a method which alleviates the problems related to leakage currents. The as-grown ferroelectric domain structure (Fig. 4(a) left) exhibits a rather fine domain structure, with an average lateral domain size around 100 nm.

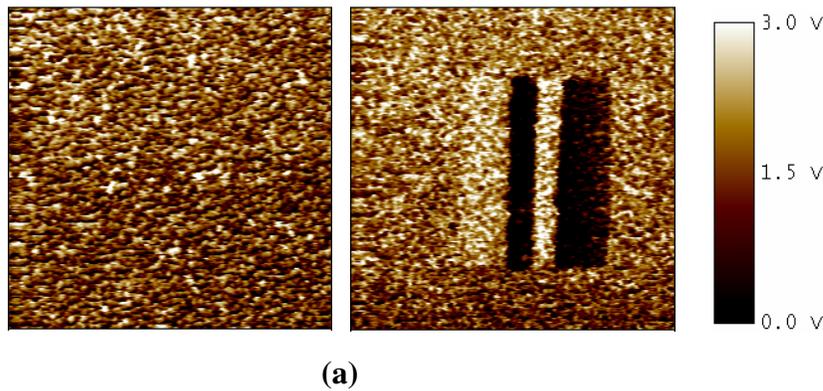

(a)

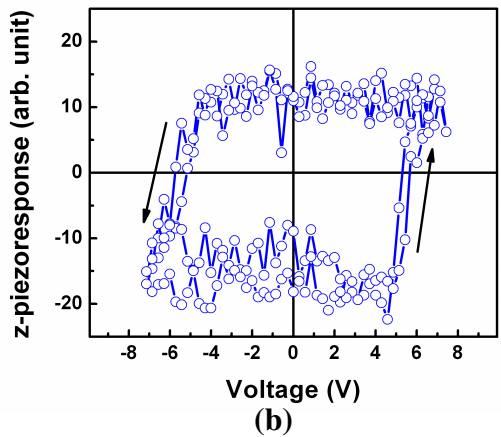

(b)

*Figure 4*. Local ferroelectric characterization of a BFCO film. **(a)** PFM image before **(left)** and after **(right)** writing oppositely polarized domains using the AFM tip as a top electrode (black/white contrast represents polarization oriented upwards/downwards). Scan size of the images is 5×5 µm². **(b)** Local remanent piezoresponse hysteresis.

To visualize the domain switching, we used the AFM tip as a movable electrode and we applied alternately positive and negative bias voltages (14 V) during a 2.5×2.5 μm$^2$ raster scan so that positively and negatively polarized stripes are created. The resulting contrast shown in Fig. 4(a) right is a clear proof that ferroelectricity in BFCO exists and polarization can be switched upon application of an external voltage. Moreover, the contrast remained unchanged for at least several hours, suggesting in addition promising retention characteristics.

A further proof of ferroelectricity, is the presence of piezoelectric hysteresis loops (remanent loop in this case), similar to the one shown in Fig. 4(b), obtained with the AFM-tip fixed above the location of the sample surface being tested. The remanent loop records the signal after a DC-bias pulse of a given voltage has been applied and switched off, therefore being insensitive to interactions of electrostatic nature.[5] The strength of the PFM signal is comparable to that obtained from BFO films, under the same measurement conditions, therefore indicating that the piezoelectric coefficients have the same order of magnitude.

Ferroelectricity of our films *at the grain level* is also demonstrated by local electromechanical measurements using PFM, as shown in Fig. 5(c). The grain encircled in Fig. 5(a) showed an initial polarization oriented downward (top to bottom, black). After a first switching of polarization by applying a positive voltage (+5V) with the AFM tip fixed in contact with the grain, the area was imaged again. Fig. 5 (b) shows that the grain exhibits opposite contrast proving that polarization has indeed been switched. The initial polarization state for this grain is restored after applying a negative voltage (-5V) (Fig. 5(c)). The hysteresis loops (not shown here) reveal that a single grain in a 250-nm thick BFCO film (about 300 nm in lateral size) switches and that the reversed polarization is stable (at the time scale of the experiment, 3 days).

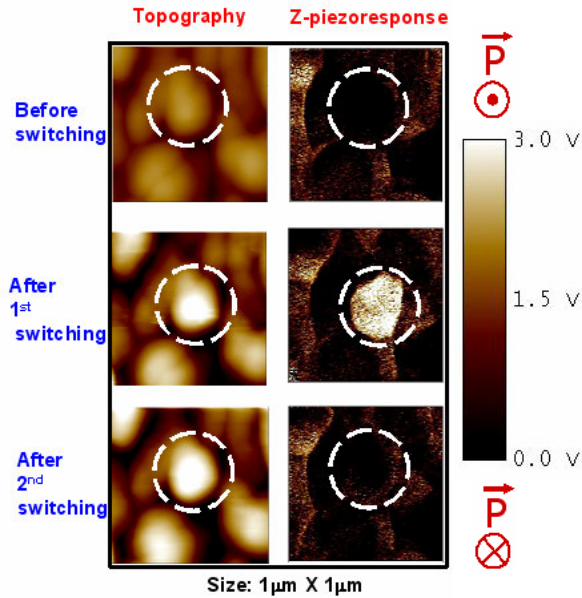

*Figure 5.* Local piezoresponse switching of a single grain of epitaxial BFCO. Scan size of the images is 1×1 μm$^2$.

## Magnetic Characterization

The presence of magnetic properties predicted by Ab-initio calculations was confirmed by magnetic measurements.

The magnetic hysteresis loop of a 300 nm thick BFCO film is compared to that of a BFO film having the same thickness in Figure 6. Both films are deposited on SRO buffer layer on STO(100). The magnetic field H was applied in the plane of the films, parallel to the [010] direction of the substrate. The saturated magnetization of BFCO was found to be ~24 emu/cc at room temperature, corresponding to ~0.31 $\mu_B$ per rhombohedral unit cell. This room temperature value is lower than the value of 2 $\mu_B$ per Fe–Cr pair predicted by ab-initio calculation for bulk BFCO at 0K, which is to be expected. These magnetic measurements clearly show a hysteretic behavior of the BFCO films with a coercive field of 80 Oe, which is about half of that measured for BFO films.

Comparison of the room temperature magnetic properties of BFCO to those of a BFO film of the same thickness [12-16] shows a higher value of the saturation magnetization for the BFCO films, which is twelve times higher than that of the BFO films. The remnant magnetization of BFCO ($2M_r$ = 2 emu/cc), however, is only showing a four-fold increase. Note that for our epitaxial BFO layers, the saturation magnetization measured was ~4 emu/cc per rhombohedral unit cell.

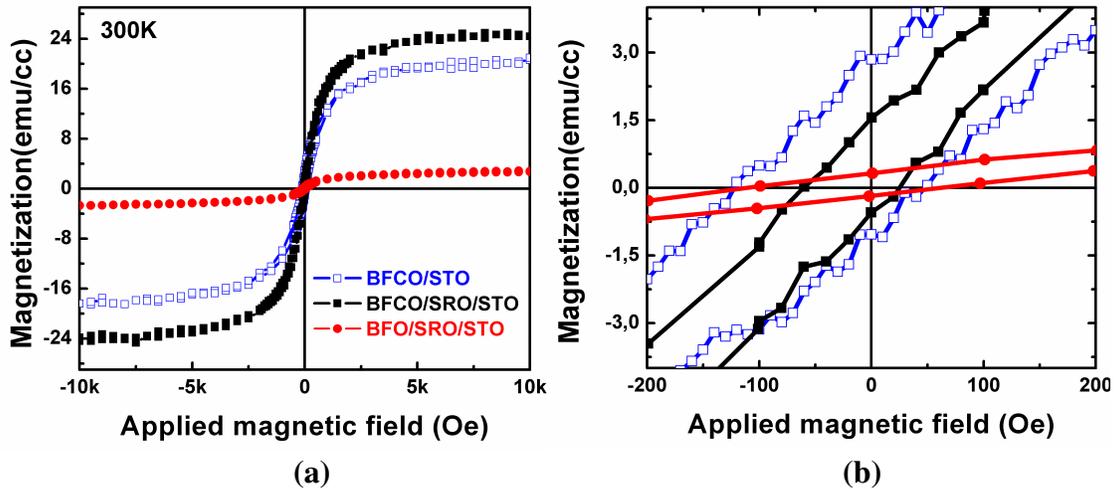

*Figure 6.* **(a)** VSM hysteresis curves at room temperature of a 300 nm thick BFCO film (squares) compared to that of a 300 nm thick BFO film (circles). The magnetic response from 285 nm thick BFCO (open squares) film deposited directly on (100)-oriented STO:Nb (without SRO electrode) is also shown. **(b)** Detailed behavior around the origin.

The presence of magnetic hysteresis at room temperature suggests that the magnetic ordering temperature is much higher than the theoretically predicted one [4] (110 K) and that the experimental magnetization values for BFCO thin films even exceed the value found very recently by Yu and Itoh in bulk BFCO ceramics, extrapolated to be around ~220K [17].

The SRO layers used as bottom electrodes and templates for the epitaxial growth of BFCO is ferromagnetic below ~160K. [11] In our measurements, done at room temperature, no direct contributions of the SRO layer to the magnetization are expected. However, a possible interface effect is observed in Figure 6, in which we compare measurements on BFCO samples grown either directly on Nb-doped STO or on an SRO layer itself grown on STO. For the film without SRO electrode, the saturation magnetization is reduced to 20 emu/cc, but the remnant magnetization and the coercive field are higher than the films with the SRO layer. Since the SRO contribution to the magnetization is indeed negligible at room temperature, which was confirmed by independent measurements (not shown here), the observed difference might be due to the effect of the different interfaces.

## CONCLUSION

In summary, we successfully prepared epitaxial BFCO films by pulsed laser deposition. The crystal structure found is very similar to that of BFO, and the films have the correct cationic stoichiometry throughout their thickness. The BFCO films are leaky but exhibit good ferroelectric and piezoelectric properties at room temperature. The insulating properties of the films are not yet fully optimized, possibly due to a slight loss of Bi during deposition. Magnetic measurements show that the films exhibit a magnetization hysteresis at room temperature with a saturation magnetization about one order of magnitude higher than that of BFO films having the same thickness. Our results partly confirm the predictions made using the ab-initio calculations about the existence of multiferroic properties in BFCO films. The existence of magnetism at room temperature is a very promising unexpected result that needs to be further investigated, and studies of the magnetic ordering and of the magnetoelectric coupling in the films are currently underway.

## ACKNOWLEDGMENTS

The authors want to thank Dr. P. Plamondon ((CM)$^2$, École Polytechnique de Montréal) for XTEM and SAED analysis and related discussions, as well as F. Normandin and Prof. T. Veres for performing preliminary magnetic measurements. Part of this research was supported by INRS start-up funds, NSERC (Canada), and FQRNT (Québec).